\documentstyle[pra,aps,amssymb,epsf]{revtex}

\newcommand{\bra}[1]{\mathop{\left\langle #1 \right|}\nolimits}
\newcommand{\ket}[1]{\mathop{\left| #1 \right\rangle}\nolimits}

\begin{document}
\draft

\title{Photoinduced Optical Rotation in a Racemic Mixture of\\
Hydrogen Peroxide Molecules}

\author{B.\ A.\ Grishanin and V.\ N.\ Zadkov\\
{\em International Laser Center and Faculty of Physics}\\{\em
M.\ V.\ Lomonosov Moscow State University, Moscow 119899, Russia}}

\maketitle

\begin{abstract}
A problem of inducing a required sign of chirality in a racemic
mixture of enantiomers of a chiral molecule is analyzed. As an
example, a racemic mixture (vapor) of left- and right-handed
enantiomers of hydrogen peroxide (H$_2$O$_2$) molecule is
considered. It is shown that biharmonic Raman excitation of the
splitted due to the left-right conversion internal rotation levels can
be effectively used for inducing optical activity in the initially
racemic vapor of H$_2$O$_2$ molecules. An experiment to study this
photoinduced optical rotation is discussed.
\end{abstract}

\widetext
\section{Introduction}
\label{intro}

Chiral specificity of the bioorganic world is one of the most
intriguing phenomenon of Nature
\cite{goldanski-ufn,holmstedt,biochir,chemchir}. It is often
thought to reflect mirror breaking responsible for a life with a
given `sign of chirality', in the absence of any evidence of
`mirror-antipode' life \cite{zeld77,hegstr80,weinberg67,salam68}.
Physical origins of this symmetry breaking are not completely
understood yet. Along with spontaneous symmetry breaking studies,
understanding that optical methods, specifically methods of
nonlinear optics \cite{akh-kor}, could be used not just to
monitor, but also to control chiral symmetry of molecules seems to
be of great importance.

\begin{figure}[thb]
\begin{center}
\epsfxsize=12cm \epsfclipon \leavevmode \epsffile{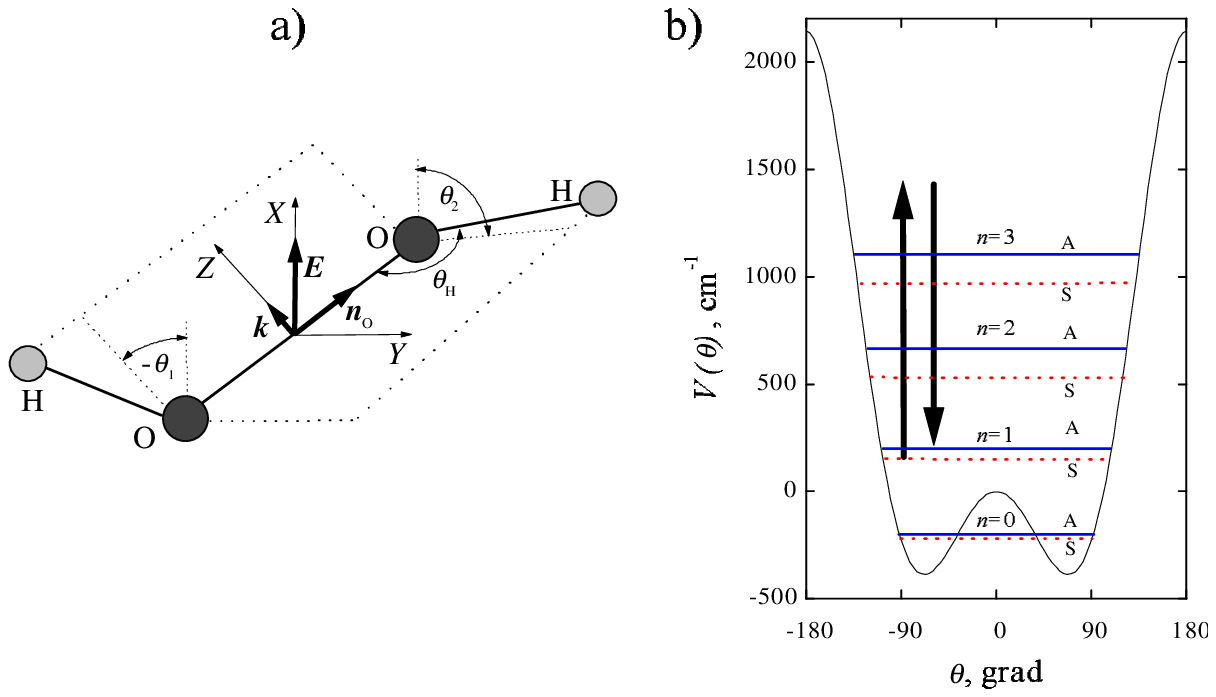}
\end{center}
\caption{a) Geometry of $\rm H_2O_2$ molecule in $d$-configuration
interacting with the incident laser field $\bf E$ wave vector $\bf
k$ of which is directed along the $Z$ axis. Equilibrium valence
angles for hydrogen atoms are equal to $\theta_{\rm H}\approx
100^\circ$, $a_{\rm O}\approx 1.461$~\AA~ and $a_{\rm
H}=0.964$~\AA~ are the O--O and O--H bond lengths,
correspondingly, ${\bf n}_{\rm O}$ is the direction vector of the
O--O bond, $\theta_1$, $\theta_2$ are the internal rotation angles
($\theta=\theta_1-\theta_2$ is the torsional angle). b) Model
torsional potential function $V(\theta)$. Pictured lower
vibrational levels  are splitted due to the tunneling through the
isomerization barrier into symmetric (dotted line) and
antisymmetric (solid line) states.} \label{fig1}
\end{figure}

In this paper we investigate a possibility of excitation of the
chiral-asymmetric states of the hydrogen peroxide molecule
(H$_2$O$_2$) driven by a laser pulse parameters of which (duration
and laser field intensity) fit special requirements discussed in
the paper. Hydrogen peroxide molecule (Fig. \ref{fig1}a) is a
simplest chiral molecule geometry of which is non-invariant with
respect to the coordinate inversion transformation
$(X,Y,Z)\to(X,Y,-Z)$.

When this transformation is applied the mirror-image of the
molecule cannot be mapped with the original with the help of any
rotation transformation because the torsional angles of the
molecule $\pm\theta=\pm(\theta_1-\theta_2)$ are not equivalent.
When we select between left- or right-handed frame, a choice of the
direction of the vector $\bf n_{\rm O}$ from one atom to another
allows us then to assign a definite sign to the torsional angle
$\angle$HOOH. An equilibrium value of the torsional angle in the
gas phase in accordance with theoretical calculations and
experimental data is $\theta\simeq\pm120^\circ$
\cite{amako62,redington62,busing65,tuylin98}, where positive sign
corresponds to the shown in Fig.~\ref{fig1}a so-called $d$-state
(right-handed enantiomer) and negative---to the $l$-state
(left-handed enantiomer).

A rough understanding of the effect of photoinduced switching
between left- and right-handed enantiomers of $\rm H_2O_2$
molecule may be obtained as follows. This molecule has a
characteristic double well torsional potential shape. Existence of
the mirror symmetry of the minima of the torsional potential leads
to the splitting of energy states as a result of tunneling through
the lower barrier. The resulting eigen states being presented with
the even and odd wave functions $\psi_S$, $\psi_A$ (tunneling
through the higher barrier and the corresponding additional
splitting are negligible \cite{tuylin98}). Relatively large
splitting value $\Delta E_0=11.4$ cm$^{-1}$ corresponds to the
absence of stationary chiral-asymmetric states, which actually
oscillate with the frequency $\omega_0=E_0/\hbar$ between the
stationary states $\psi_S,$ $\psi_A$ until the symmetric
equilibrium is reached. As a result, one can manipulate the chiral
properties of $\rm H_2O_2$ molecule inducing tunneling between
$d$- and $l$-enantiomers and therefore the effect of photoinduced
optical rotation has an oscillating character. By contrast, for
the case of heavier molecules tunneling time between different
enantiomers of which may be infinitely large and initially produced
chiral asymmetric state is stable.

The Hamiltonian of molecule--laser field interaction $\hat H_I=
-\sum{\bf d}_k {\bf E}({\bf r}_k)$ displays a qualitative
difference in dipole $(H_D)$ and quadrupole $(H_Q)$
approximations. In dipole approximation (at ${\bf E}({\bf r}_k)\to
{\bf E}({\bf r}_0)$) proton contributions are added, whereas in
quadrupole approximation (at ${\bf E}({\bf r}_k)\to ({\bf
r}_k-{\bf r}_0)\nabla\,{\bf E}({\bf r}_0)$) they are substracted.
As a result, $H_D$ is an even function of the torsional angle and
$H_Q$ is an odd one. For the corresponding off-diagonal matrix
elements $H^{12}_D$, $H^{12}_Q$ of the $S\to A$ transition we get
\begin{equation}\label{H12}
H^{12}_D=0,\quad H^{12}_Q\neq0.
\end{equation}

\noindent This means that the precession between the tunneling
splitted right and left chiral states ($d\leftrightarrow l$) can
be excited only due to the quadrupole interaction, whereas the
dipole interaction causes only modulation of the eigen states
energies, oscillating with the frequency of the exciting laser
field. In other words, we can {\em selectively} excite due to the
quadrupole interaction $d$- or $l$-enantiomers using one of the
most fundamental quantum optics effects---coherent precession of a
two-level system under the action of a coherent pulse of
electromagnetic field.

In this paper we discuss the conditions under which the effect of
photo-induced chirality can be studied experimentally by
registering the corresponding photoinduced optical activity in a
racemic mixture of left- and right-handed enantiomers of the
H$_2$O$_2$ molecule, which shows no optical activity otherwise.

A rough estimate of the order of magnitude of the discussed effect
can be made within the model of one-photon excitation. The
criterion of alignment of the molecules dipole moments along $Z$
axis for 100\% of the molecules in a solution under the action of
dc electric field with the magnitude $E$ can be written as
$Eea_{\rm H}\gg kT$ ($a_{\rm H}$ is the O--H bond length, $e$ is
the proton charge), which at the room temperature gives
$E\gg10^6$~V/cm. If we do not require that 100\% of the molecules
in the solution are to be aligned, we can then use relatively
moderate field intensities and therefore achieve a smaller degree
of alignment $\varkappa=Eea_{\rm H}/kT$, which however may be
enough to determine experimentally optical activity in the
initially racemic solution.

In the frequency region of electronic susceptibility, the effect
can be estimated to the lowest order under the suggestion that
$\varkappa$ represents the part of molecules aligned strictly
along the field, whereas the rest of the molecules do not
contribute to the effect. Then the upper estimate can be obtained
by multiplying the typical specific rotation value
$\alpha\sim10^2$ deg$\cdot$cm$^3$/g$\cdot$dm for the media with
strong optical activity by the small parameter $\varkappa$, which
gives us finally $\sim10^{-1}$ deg$\cdot$cm$^3$/g$\cdot$dm for the
resonantly excited optical activity at $E\approx10^3$~V/cm. For a
vapor at normal conditions and at the interaction length of 1~dm
the rotation angle is about $0.1''$. This value is about the
sensitivity limit of the linear polarization spectroscopy methods.
Therefore, even without going into other obstacles, it seems
difficult to observe the effect of photoinduced optical rotation
in the scheme with one-photon resonant excitation and using dc
electric field for aligning the molecules.

Additional trouble is the finite lifetime $\tau_r$ of the chiral
state due to the molecular collisions in a solution. This time
should be at least longer than the time the laser field travels
through the active region $\tau_c= L/c \ge 10^{-10}$~s$^{-1}$ and
its lower bound estimate  is given by $\tau_r=({\cal
N}v\sigma)^{-1}$, where ${\cal N}$ is the molecular concentration,
$v$ is the heat velocity and $\sigma$ is the collisional cross
section. At the normal conditions one gets $\tau_r\sim10^{-9}$
s$^{-1}> \tau_c$.

Two-photon Raman excitation scheme we discuss in this paper fits
the above condition and, beyond that, provides an effective
excitation of the required chiral state of the molecule using
lasers in the convenient frequency range.

\section{H$_2$O$_2$ molecule's dynamics in electromagnetic field}

\subsection{Reduced model of free molecule's dynamics}
\label{:reduc}

To describe free molecule's dynamics driven by the laser field we
shall start with the mathematical description of free molecule's
dynamics. For this it seems helpful to reduce the total molecular
Hamiltonian using the fact that proton to oxygen mass ratio is a
small one. Then the protons' dynamics can be treated adiabatically
with respect to the oxygen coordinates so that the dynamics of
oxygen nuclei can be calculated by averaging over direction of the
unit vector ${\bf n}_{\rm O}=(\sin\vartheta \cos \varphi,
\sin\vartheta\sin\varphi, \cos\vartheta)$ along the O--O bond.
Averaging along $Z$-coordinate (Fig. \ref{fig1}a) should be
performed then only for protons rotation about the O--O bond,
which potential depends only on the torsion angle $\theta=
\theta_2 - \theta_1$.

Under these conditions, the molecular Hamiltonian in a reduced
model, which includes both the free molecule's rotation about the
O--O bond and torsional oscillations, takes the form:
\begin{equation}
\label{HH} \hat H=\hat H_{\rm H}+\hat H_\theta
\end{equation}

\noindent with
\begin{eqnarray}
\label{HHH} \hat H_{\rm H}&=&-\frac{\hbar^2}{4m_{\rm H}r_{\rm
H}^2} \frac{\partial^2}{\partial\tilde\theta^2},\\ \label{HHO}
\hat H_\theta&=&-\frac{\hbar^2}{m_{\rm H}r_{\rm H}^2}
\frac{\partial^2}{\partial\theta^2}+V(\theta),
\end{eqnarray}

\noindent where $\tilde\theta=(\theta_1+\theta_2)/2$, $m_{\rm H}$
is the proton mass and $V(\theta)$ is the torsional potential. For
simplicity, we neglect the vibrations of the valence angles
$\angle$HOO.

Initial molecule's position, which can be referred to as the mean
angle $\tilde\theta$, includes, in addition to the above mentioned
uncertainty in direction ${\bf n}_{\rm O}$, an uncertainty due to
the molecule's rotation about this direction, which corresponds to
the third rotational degree of freedom plus to the two angles that
determine the direction of ${\bf n}_{\rm O}$. Corresponding to the
free rotation about {O--O} bond transition frequencies
$\omega_{n\to n+1}= 7.84\times (n+1/2)$~cm$^{-1}$ calculated from
(\ref{HHH}), (\ref{HHO}) are four times smaller than the
corresponding frequencies of internal rotation, which are
determined by the torsional potential $V(\theta)$.

\subsection{Photoexcitation of molecule's rotational degrees of freedom}

In this section we will discuss a model of molecule--laser field
interaction with an accent placed on the photoexcitation of
molecule's rotational degrees of freedom. Let us consider
H$_2$O$_2$ molecule interacting with the laser field ${\bf
E}(t)={\bf E}_1u_1(t) \cos(\omega_1t+ \varphi_1)$ with the pulse
envelope $u_1(t)$ and frequency $\omega_1\gg\omega_ {n\to n+1}$, so
that we can neglect the quantum specificity of the excitation. The
following classical equation governs then the rotational dynamics
of the molecule: $$
J\frac{d^2\tilde\theta}{dt^2}=\frac{\partial}{\partial\tilde
\theta}{\bf E}(t){\bf d}, $$

\noindent where $J$ is the molecule's moment of inertia.

In zero'th order with respect to the angle deviation for the
acting force in the right side of the equation and in the case of
not too short laser pulses ($\tau_1 \gg1/\omega_1$) the angle
response $\Delta\tilde\theta$ at the frequency $\omega_1$ takes
the form:

\begin{equation}\label{dtheta}
\Delta\tilde\theta(\omega_1)\simeq-\frac{\partial}{\partial\tilde\theta}
\frac{{\bf E}_1(t){\bf d}}{J\omega_1^2}.
\end{equation}

\noindent This angle deviation causes the modulation of the
quadrupole Hamiltonian at the frequency $\omega_1$, which in its
turn causes the internal rotation of the molecule and therefore
leads to the partial aligning of the molecule. The latter reveals
in modification of the quadrupole Hamiltonian rotational symmetry.
In particular, in case the O--O bond directs along Z axis, i.e.
$\vartheta=0$, and the laser field ${\bf E}_1$ shines along Y
axis, the rotational symmetry leads to a homogeneous distribution
versus angle $\tilde\theta$, dependence on which in the quadrupole
moment is given by $\sin \tilde\theta$. Molecule's orientation
reveals then in the angle response, which can be extracted from
Eq.\ (\ref{dtheta}) in the form
\begin{equation}\label{dthetas}
\Delta\tilde\theta\simeq\frac{E_1d}{2J\omega_1^2}u_1(t)\cos
\tilde\theta\cos\frac{\theta}{2}\cos(\omega_1t+\varphi_1),
\end{equation}

\noindent where the angle dependence $\cos\tilde\theta$ follows
that one of the quadrupole Hamiltonian for the laser field ${\bf
E}_2\|X$, so that
$$
\Delta\hat H_Q=\partial\hat H_Q/\partial
\tilde\theta\, \Delta\tilde\theta \propto\cos^2\tilde\theta.
$$

\noindent Estimation of the order of magnitude for the visual
range of optical frequencies is given by
\begin{equation}\label{dthnum}
\Delta\tilde \theta\sim10^{-10}\sqrt{I_1},
\end{equation}

\noindent where $I_1$ is the ${\bf E}_1$ field intensity in
W/cm$^2$ and the angle is given in radians. This estimate goes to
1 only for the field strengths larger than atomic ones.

Torsional oscillations dynamics describing by the Hamiltonian
$\hat H_\theta$ is essentially quantum one due to the relatively
small proton to oxygen mass ratio. This leads to the tunneling
between the $d$ and $l$ local minima of the torsional potential
(Fig. \ref{fig1}b) and therefore to two energy-splitted
superposition eigen states $\psi_S$, $\psi_A$ with equally
represented $d$- and $l$-configurations. In addition, it
determines the non-rigidity of molecule's geometry due to the
quantum uncertainty of the wave functions versus torsional angle.
Torsional potential of H$_2$O$_2$ molecule and its eigen energies
have been extensively explored by {\em ab initio} calculations and
respective fitting to the experimental spectra. The torsional
potential and corresponding energy levels structure taken for our
calculations from \cite{tuylin98} are shown in Fig. \ref{fig1}b.
With this potential, an estimate for the local states uncertainty
on torsional angle, which is based on expression for the
fluctuations in vacuum state of a harmonic oscillator, gives
$\sigma_\theta\approx[\hbar/ (m_Hr_H^2\omega_0)]^{1/2}
\approx20^\circ$.

The interaction Hamiltonian in dipole approximation, which takes
into account only proton charges displacements, has the form:
\begin{equation}\label{HD}
\hat H_D=-E_Lea_{\rm H}\Re{\bf e}({\bf e}_1+{\bf e}_2),
\end{equation}

\noindent where ${\bf e}_{1,2}$ are the corresponding unit
direction vectors of the protons bonds, ${\bf e}$ is the laser
field polarization vector and $e$ is the proton charge. The
quadrupole contribution to the interaction Hamiltonian depends on
the choice of the center of frame. Its displacement causes an
additional contribution to the dipole Hamiltonian, which is
however small and can be neglected as it displays the same
qualitative interaction properties as the main term. In our
calculations we therefore choose the center of frame as it is
shown in Fig.~\ref{fig1}a in order to simplify the Hamiltonian.
The corresponding expression takes the form:
\begin{equation}\label{HQ}
\hat H_Q=-{\bf E}_L{\bf d}_Q=-\frac{{\bf k}{\bf n}_{\rm O} a_{\rm
O}}{2}E_Lea_{\rm H}\Re{e}\, i{\bf e}({\bf e}_2-{\bf e}_1),
\end{equation}

\noindent where $a_{\rm O}$ is the O--O bond length. The discussed
choice of the center of frame leads to the dependence of
quadrupole Hamiltonian $H_Q$ on torsional angle given by the odd
function $\sin\theta/2$.

To proceed with calculations of Eqs (\ref{HD}), (\ref{HQ}) the
polarization vector should be represented in the form ${\bf
e}=(e_X,e_Y,0)$ with the components $e_X$, $e_Y$, which are
generally the complex numbers. Then the corresponding coefficients
$$
C_{H_k}={\bf n}_X \cdot {\bf n}_{H_k},\qquad S_{H_k}={\bf n}_Y
\cdot {\bf n}_{H_k}
$$

\noindent can be calculated analytically. Here ${\bf n}_X$, ${\bf
n}_Y$ are the unit vectors along axes $X$, $Y$ and ${\bf n}_{H_k}$
are the unit vectors along the bonds O--H$_k$, $k=1,2$. Under
these notations Hamiltonians (\ref{HD}), (\ref{HQ}) take the form
\begin{equation}
\label{HDQ}
\begin{array}{lcl}
\hat H_D&=&-E_Lea_{\rm H}\Re{e}\,
\displaystyle e^{-i\varphi_L}[e_X(C_{H_1}+C_{H_2})+e_Y(S_{H_1}+S_{H_2})],\\
\hat H_Q&=&-\displaystyle\frac{k_La_{\rm O}}{2}E_Lea_{\rm H} \Re{e}\,
ie^{-i\varphi_L}[e_X(C_{H_2}-C_{H_1})+e_Y(S_{H_2}-S_{H_1})],
\end{array}
\end{equation}

\noindent where $k_L$ is the modulus of the laser field wave
vector, $\varphi_L$ is the laser field phase, and rotation angles
$\theta_{1,2}$ of proton bonds represent coordinate operators. The
projection coefficients of the proton dipole moments onto the
directions $X$, $Y$ of the polarization vector ${\bf e}$ are given
by the following formulae
\begin{eqnarray*}
C_{H_k}&=&{\bf n}_X\,O({\bf n}^\bot_{OOH_1},\delta_H) O({\bf
n}_{\rm O},\theta_k)\,O({\bf n}^\bot_{{\bf n}_X{\bf n}_O},\pi/2-
\angle\,{{\bf n}_X{\bf n}_{\rm O}})\,{\bf n}_X,\\
S_{H_k}&=&{\bf n}_Y\,O({\bf n}^\bot_{OOH_2},\delta_H) O({\bf n}_{\rm
O},\theta_k)\,O({\bf n}^\bot_{{\bf n}_X{\bf n}_O},\pi/2-
\angle\,{{\bf n}_X{\bf n}_{\rm O}})\,{\bf n}_X
\end{eqnarray*}

\noindent as the scalar product of vector ${\bf n}_{X,Y}$ and
another vector obtained as follows. First, the vector ${\bf n}_X$
is rotated at the angle $\pi/2-\angle\,{{\bf n}_X {\bf n}_{\rm
O}}$ in the plain ${\bf n}_X{\bf n}_{\rm O}$ until it is
perpendicular to the axis ${\bf n}_{\rm O}$ which is used as a
reference axis torsional angles are calculated in respect to
which. Second, the vector ${\bf n}_X$ is rotated about axis O---O
at the torsional angle $\theta_k$. Third, the subsequent rotation
of ${\bf n}_X$ vector is performed in the plane OOH$_k$ at the
angle $\delta_H= \theta_H-\pi/2$ until we get the unit vector
along the vector ${\bf n}_{\rm H_k}$.

When non-zero value $\delta_H$ is taken into account, analytical
expressions for the above listed coefficients derived with the use
of computer algebra are bulky, so that we list them here for a
specific case of $\delta_H=0$, i.e., for the case when the proton
bonds are orthogonal to the bond O--O. The corresponding error is
then not larger than $\sim10$\% due to the small value of
$\delta_H  \approx 10^\circ$. The matrix of three-dimensional
rotations at angle $\alpha$ about axis $\bf n$ is given then by
$$
O({\bf n},\alpha)\!=\! \left(\begin{array}{ccc} n_x^2 \!+\!
n_y^2\cos\alpha \!+\! n_z^2\cos\alpha&
   n_xn_y \!-\! n_xn_y\cos\alpha \!-\! n_z\sin\alpha&
   n_xn_z \!-\! n_xn_z\cos\alpha \!+\! n_y\sin\alpha\\
   n_xn_y \!-\! n_xn_y\cos\alpha \!+\! n_z\sin\alpha&
   n_y^2 \!+\! n_x^2\cos\alpha \!+\! n_z^2\cos\alpha&
   n_yn_z \!-\! n_yn_z\cos\alpha \!-\! n_x\sin\alpha\\
   n_xn_z \!-\! n_xn_z\cos\alpha \!-\! n_y\sin\alpha&
   n_yn_z \!-\! n_yn_z\cos\alpha \!+\! n_x\sin\alpha&
   n_z^2 \!+\! n_x^2\cos\alpha \!+\! n_y^2\cos\alpha
\end{array}\right).
$$

\noindent With the use of this expression the
resulting coefficients takes the form:
\begin{equation}
\label{CHSH}
\begin{array}{lcl}
&&C_{H_1}=\sqrt{\displaystyle1\-\sin^2\vartheta\cos^2\varphi}\cos\theta_1\;,\quad
C_{H_2}=\sqrt{\displaystyle1-\sin^2\vartheta\cos^2\varphi}\cos\theta_2\;,\\
S_{H_1}&=&\displaystyle\left(16\sqrt{\displaystyle\cos^2\vartheta+\sin^2\varphi\sin^2\vartheta}
\right)^{-1}\left[-2\sin(2\varphi-\theta_1)+\sin(2\varphi-2\vartheta-\theta_1)
-8\sin(\vartheta-\theta_1)+\right.\\
&&+\left.\sin(2\varphi+2\vartheta-\theta_1)-
2\sin(2\varphi+\theta_1)+\sin(2\varphi-2\vartheta+\theta_1)
+8\sin(\vartheta+\theta_1)+\sin(2\varphi+2\vartheta+\theta_1)\right]\;,\\
S_{H_2}&=&\displaystyle\left(16\sqrt{\cos^2\vartheta+\sin^2\varphi\sin^2\vartheta}\right)^{-1}
\left[-2\sin(2\varphi-\theta_2)+\sin(2\varphi-2\vartheta-\theta_2)
-8\sin(\vartheta-\theta_2)+\right.\\
&&+\left.\sin(2\varphi+2\vartheta-\theta_2)-2\sin(2\varphi+\theta_2)+
\sin(2\varphi-2\vartheta+\theta_2)+8
\sin(\vartheta+\theta_2)+\sin(2\varphi+2\vartheta+\theta_2)\right].
\end{array}
\end{equation}

\noindent For the sum of the coefficients determining the dipole
interaction potential we get then the following expressions
\begin{equation}
\label{cH2PcH1}
\begin{array}{lcl}
&&C_{H_1}+C_{H_2}=\displaystyle C_+\cos\frac{\theta}{2}\,,\quad
S_{H_1}+S_{H_2}=S_+\cos\frac{\theta}{2}\,,\quad
C_+=2\sqrt{1-\sin^2\vartheta\cos^2\varphi}\,\cos\tilde\theta\,,\\
S_+&=&2\displaystyle\left(16\sqrt{\cos^2\vartheta+2\sin^2\varphi\sin^2\vartheta}\right)^{-1}
[-2\sin(2\varphi-\tilde\theta)-2\sin(2\varphi+\tilde\theta)+
\sin(2\varphi-\tilde\theta-2\vartheta)+\\
&&+\sin(2\varphi+\tilde\theta-2\vartheta)+8\sin(\tilde\theta-\vartheta)+
8\sin(\tilde\theta+\vartheta)+\sin(2\varphi-\tilde\theta+
2\vartheta)+\sin(2\varphi+\tilde\theta+2\vartheta)],
\end{array}
\end{equation}

\noindent which depend on the torsional angle as the even function
$\cos\theta/2$. For the difference of the coefficients determining
quadrupole interaction potential we get the following expressions,
which depend on the torsional angle as the odd function
$\sin\theta/2$:
\begin{equation}
\label{sH2MsH1}
\begin{array}{lcl}
&&C_{H_2}-C_{H_1}=C_-\sin\displaystyle\frac{\theta}{2}\,,\qquad
S_{H_2}-S_{H_1}=S_-\sin\frac{\theta}{2}\,,\quad
C_-=-2\displaystyle\sqrt{1-\sin^2\vartheta\cos^2\varphi}\,\sin\tilde\theta\,,\\
S_-&=&\displaystyle\left(16\sqrt{\cos^2\vartheta+2\sin^2\varphi\sin^2\vartheta}
\right)^{-1}[(2\cos(2\varphi-\tilde\theta)-2\cos(2\varphi+\tilde\theta)-
\cos(2\varphi-\tilde\theta-2\vartheta)+\\
&&\displaystyle+\cos(2\varphi+\tilde\theta-2\vartheta)+8\cos(\tilde\theta-\vartheta)+8
\cos(\tilde\theta+\vartheta)-\cos(2\varphi-\tilde\theta+2\vartheta)
+\cos(2\varphi+\tilde\theta+2\vartheta)].
\end{array}
\end{equation}

For the simplified analysis of both molecule's rotation about axis
O--O and torsional oscillations, the numerical values of the
torsional potential $V(\theta)$ are essential. The corresponding
numbers for the energy splitting, which we adopted from
\cite{tuylin98}, are as follows: $\Delta E_0=11.44$~cm$^{-1}$,
$\Delta E_1= 116.34$~cm$^{-1}$, $\Delta E_2=206.57$~cm$^{-1}$. For
$n\ge1$ they exceed the frequencies of free molecule rotation.
Therefore, the transitions between the eigen states of the
torsional Hamiltonian $\hat H_\theta$ can be described in terms of
classical coordinate $\tilde\theta$.

\section{Photoinduced internal rotation in H$_2$O$_2$ molecule}
\label{:TLM}

This section presents a discussion on photoinduced excitation of
internal conversion in H$_2$O$_2$ between left- and right-handed
enantiomers.

Torsional potential of H$_2$O$_2$ molecule has a characteristic
double-well shape (Fig.\ 1b) minima of which correspond to the
left- and right-handed enantiomers and are mirror-symmetrically
spaced. As a result, the eigenstates of the torsional Hamiltonian
split due to the tunneling through the potential energy barrier
separating the two wells. It seems reasonable to assume that the
splitting of the eigen states of the torsional Hamiltonian $\hat
H_\theta$, or internal rotation frequencies, significantly exceed
the rotation frequencies of a free molecule. Under this
assumption, resonant laser excitation of the internal rotation
transition causes negligibly small deviations in molecule's
orientation angles $\vartheta,$ $\varphi$ and a small deviation of
the rotation angle $\tilde \theta$. Therefore, the symmetric,
$\psi_S$, and antisymmetric, $\psi_A$, states of the torsional
potential can be excited into a coherent superposition $C_A\psi_A
+ C_S\psi_S$. The transition energy between $S$- and $A$-states,
$\Delta E_n$, can be estimated as $$ \Delta E\sim \exp
\left(-2\sqrt{m_{\rm H}a_{\rm H}^2\Delta V} \Delta\theta/
\hbar\right), $$

\noindent where $\Delta V$ and $\Delta \theta$ are the
characteristic values of the height and width of the potential
barrier (Fig.\ 1b).

In case of resonant excitation with the frequency $\omega_0=\Delta
E_n/\hbar$ only resonant matrix elements in the Hamiltonians
(\ref{HDQ}) remain essential and, therefore, torsional dynamics of
the molecule can be essentially reduced to that one of the
two-level system.

Let us now examine, for the case of linear polarization ($e_X=1$,
$e_Y=0$)\footnote{Using circular polarization gives no advantage
in quadrupole approximation because polarization does not affect
the dependence of the quadrupole moment of the transition on
coordinate $\theta$.}, the 2$\times$2-matrix of the one-photon
interaction Hamiltonian $\hat H_L$ of the total one-photon
interaction Hamiltonian $\hat H_I=\hat H_D+ \hat H_Q$, where with
the help of Eqs (\ref{HDQ}) $\hat H_D\propto\cos\theta$ and $\hat
H_Q \propto \sin\theta.$ The total interaction Hamiltonian $\hat
H_I$ can be written using Pauli matrices of the transition in the
form:
$$
\hat H_I\to V_{12}\hat\sigma^++V_{21}\hat\sigma^-=
V_{12}(t)\hat\sigma_1(t)=V_{12}(0)\cos(\omega_Lt+ \tilde\varphi_L)
(\cos\omega_Lt\,\hat\sigma_1+ \sin\omega_Lt \,\hat\sigma_2),
$$

\noindent where $V_{12}=V_{21}$ due to the fact that the eigen
functions $\psi_k$ are real-valued.

In the rotation wave approximation (RWA) (see, for example,
\cite{allenr}), after averaging over oscillations of the field and
transition polarization at the frequency $2\omega_L$ we get with
the help of Eqs (\ref{HDQ}), (\ref{cH2PcH1}), (\ref{sH2MsH1})
\begin{equation}
\label{HL}
 \hat H_L=
  \left(\begin{array}{cc}
    0 &\displaystyle \frac{1}{2}QS_{AS}e^{-i\tilde\varphi_L}\\
    \displaystyle\frac{1}{2}QS_{AS}e^{i\tilde\varphi_L} & 0
  \end{array}\right),
\end{equation}

\noindent where
\begin{equation}
\label{HLTLM} Q=k_La_{\rm O}E_Lea_{\rm H}
\sqrt{1-\sin^2\vartheta\cos^2\varphi}\,\sin\tilde\theta,
\end{equation}

\noindent $\tilde\varphi_L$ is the initial phase of laser field,
including phase contribution, which is determined by the field
polarization and molecule orientation in accordance with Eq.\
(\ref{HDQ});
\begin{equation}
\label{cs}
 S_{AS}=\int\limits_{-\pi}^\pi\psi_A(\theta)
\sin\frac{\theta}{2}\,\psi_S(\theta)\,d\theta
\end{equation}

\noindent is the dimensionless matrix element of the tunneling
transition.

We can re-write the Hamiltonian (\ref{HL}) in the matrix form
\begin{equation}
\label{HRWA}
 \hat H_\Omega=\left(
  \begin{array}{cc}
    -\displaystyle\frac{\hbar\delta}{2} &
    \displaystyle\frac{Q}{2}S_{AS}e^{-i\tilde\varphi_L} \\
    \displaystyle\frac{Q}{2}S_{AS}e^{i\tilde\varphi_L} &
    \displaystyle\frac{\hbar\delta}{2}
  \end{array}\right),
\end{equation}

\noindent where $\delta=\omega_L-\omega_{12}$ is the laser field
detuning. This operator can also be expressed with the use of
Pauli matrices in the form
\begin{equation}\label{vecO}
\hat H_\Omega=\displaystyle\frac{\hbar}{2}
\vec\Omega\cdot\hat{\vec\sigma},\quad \vec\Omega=(-\delta,
QS_{AS}\cos\tilde\varphi_L,\sin\tilde\varphi_L).
\end{equation}

\noindent Time evolution operator
$$ U(t)={\rm
T}\,\exp[(-i/\hbar)\int \hat H_\Omega\,dt] $$

\noindent corresponding to the operator (\ref{HRWA}) can be
calculated analytically for the case of a) rectangular pulse $E_L
={\rm const}$ or b) zero detuning $\delta=0$ using the following
formulae:
\begin{equation}
\label{Ut}
\begin{array}{lcl}
\mbox{a)}\quad U(t)&=&\cos\left(\displaystyle \frac{\Omega}{2}t\right)\hat
I-i\sin\left(\displaystyle \frac{\Omega}{2}t\right)\left[-\displaystyle
\frac{\delta}{\Omega}\hat\sigma_3+\frac{\displaystyle QS_{AS}}{\Omega}
(\hat\sigma_1\cos\tilde\varphi_L+\hat\sigma_2\sin\tilde\varphi_L)\right],\\
\mbox{b)}\quad U(t)&=&\cos\left( \displaystyle\frac{\Phi}{2}\right)\hat I-
i\sin\left( \displaystyle\frac{\Phi}{2}\right)
(\hat\sigma_1\cos\tilde\varphi_L+\hat\sigma_2\sin\tilde\varphi_L),
\end{array}
\end{equation}

\noindent where $\Omega=\sqrt{\Omega_0^2+\delta^2}$ is the total
Rabi frequency, $\Omega_0=QS_{AS}$ is the standard Rabi frequency,
and $\Phi= \int\Omega(t)\,dt$ is the laser pulse angle.

For $\delta=0$, we immediately find from (\ref{Ut}b) that initial
incoherent states represented as $\hat\rho_0=\hat I/2+w\hat
\sigma_3/2$ ($-1\leq w\leq 1$) transform into the states
\begin{equation}
\label{rhot}
\hat\rho_t=\frac{1}{2}\left[\hat I+w(\cos\Phi\,\hat\sigma_3 -
\sin\Phi\sin(\omega_Lt+\tilde\varphi_L)\hat\sigma_1-
\sin\Phi\cos(\omega_Lt+\tilde\varphi_L)\hat\sigma_2)\right].
\end{equation}

\noindent Here, in addition to the transformation (\ref{Ut}), free
precession with the laser field frequency is also taken into
account (by contrast, the latter is regularly applied in the
interaction representation \cite{landauqm} and RWA to the
operators of physical variables). In (\ref{rhot}) the terms with
$\hat\sigma_1$, $\hat\sigma_2$ represent the contribution of
coherent superposition of the states $\psi_S$, $\psi_A$. In
particular, for the lower initial state, which corresponds to
$w=1$ at $\omega_Lt+\tilde\varphi_L=\pi/2$, excitation of the
system by $\pi/2$-pulse, for which $\Phi=\pm\pi/2$, $\cos\Phi=\pm
1$ in (\ref{rhot}) and $\Phi/2= \pm\pi/4$, $\cos(\Phi/2)=\pm
1/\sqrt{2}$ in (\ref{Ut}b), transfers the initial state $\psi_S$
into the chiral states $\psi_{1,2}= (\psi_S\pm\psi_A) /\sqrt{2}$
corresponding to the density matrices $(\hat I\mp\hat\sigma_1)/2$.
Therefore, for the fixed angles $\Theta=(\vartheta, \varphi,
\tilde\theta)$ it is possible to switch the molecule into $d$- or
$l$-state with 100\% probability exciting tunneling transition by
a laser pulse with properly adjusted parameters. In a more general
case, however, the resulting state is to be averaged over the
angles $\Theta$.

This averaging can be done with the use of standard superoperator
calculation technique \cite{gQEDr}. Performing averaging for the
resulting excitation (\ref{rhot}) of the initially incoherent
state just over the sign of the parameter $Q$, which depends on
the angle $\tilde\theta$ of the mean proton bonds orientation
along axis $X$ that is accounted in (\ref{rhot}) by factor
$\sin\Phi$, we receive the following structure of the density
matrix:
$$
\hat\rho_t=(\hat I-w\cos\Phi\hat\sigma_3)/2.
$$

\noindent This means that in the absence of molecules orientation
over angle $\tilde\theta$ the density matrix is transformed
incoherently, i.e., diagonal matrices preserve the diagonal form.
Therefore, after such kind of transformation the symmetry of each
pure component $S$ or $A$ does not change so that the square
modulus of the wave function preserves mirror symmetry with
respect to the transformation $\theta \to -\theta$. As a result,
we can conclude that averaging over the $\Theta$ angles cancels
the effect.

How can we cope with this? An answer is that to broke the mirror
symmetry the molecules should be spatially aligned during the
excitation. In the section below we will examine the excitation
of preliminary aligned molecules.

\subsection{Excitation of preliminary aligned molecules}
\label{:orient}

Let us assume that we were able spatially align the molecules in
an ensemble. This can be done either by applying strong dc
electric field (see Introduction) or by a laser pulse (sections 2,
4). If the molecules are spatially aligned, an inhomogeneity in
the distribution of angle $\tilde\theta$ takes place and,
therefore, excitation of incoherent states will contain a coherent
component corresponding to the excitation of chiral states, i.e.,
the ones different from $\psi_A$ and $\psi_S$. It is desirable to
be able to characterize the effect of excitation by a simple
characteristic. Let us determine the scalar property of the degree
of chirality as the average value
\begin{equation}
\label{chi}
\chi=2\left(\bra{\psi_l}\hat\rho\ket{\psi_l}-\frac{1}{2}\right)=
-2\left(\bra{\psi_d}\hat\rho\ket{\psi_d}-\frac{1}{2}\right),
\end{equation}

\noindent where
$$
\psi_{l,d}=\frac{1}{2}[\psi_S\pm\psi_A]
$$

\noindent represents left and right chiral states,
correspondingly, for which the corresponding degrees of chirality
are $\chi=\pm1$ for the pure states $\hat\rho=\ket{\psi_{l,d}}
\bra{\psi_{l,d}}$. For the state $\hat\rho(t)$ excited by a
rectangular laser pulse with duration $\tau_p$ and pulse angle
$\Phi=\Omega\tau_p$ the dependence of the degree of chirality on
detuning $\delta$, angle $\Phi$, and phase $\tilde\varphi_L$ can
be calculated analytically with the use of computer algebra.

In the simplest case, for zero laser detuning, this dependence
takes the form
$$
\chi=-\sin\Phi\sin\tilde \varphi_L.
$$

\begin{figure}[tbh]
\begin{center}
\epsfxsize=15cm \epsfclipon \leavevmode \epsffile{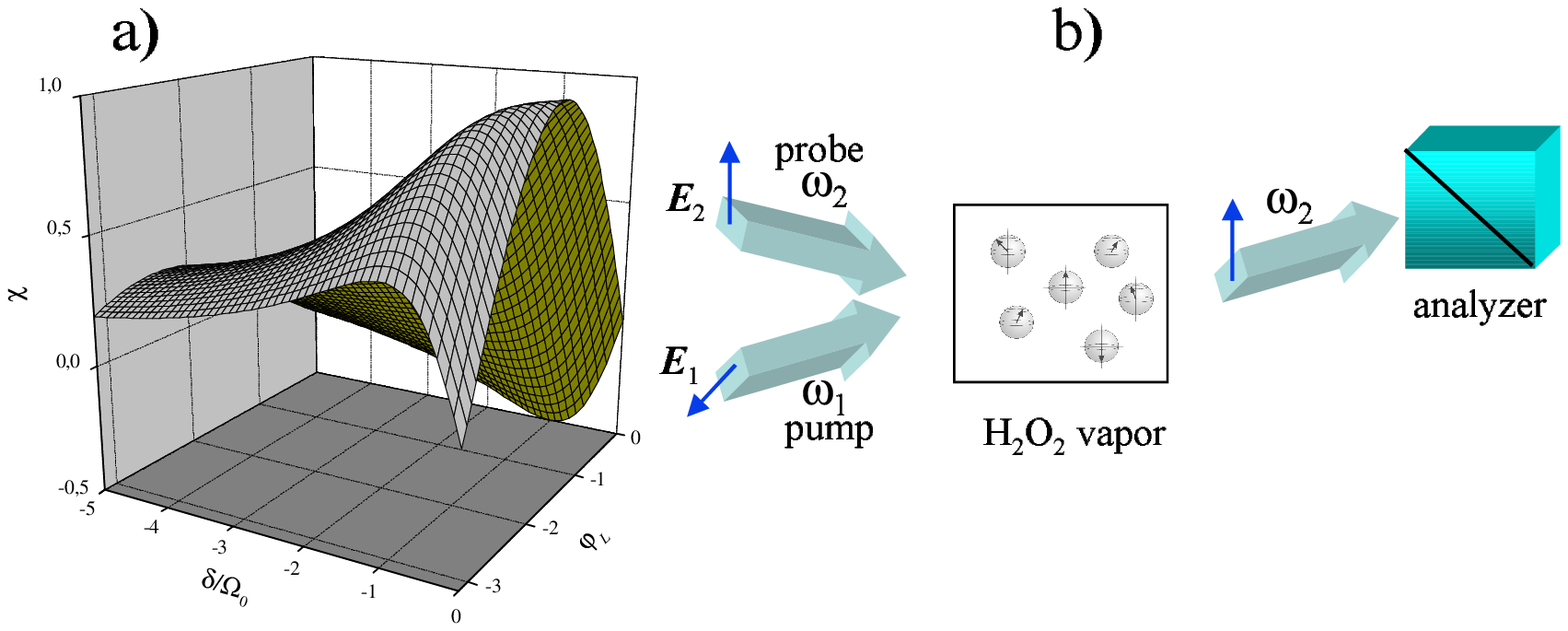}
\end{center}
\caption{a) Degree of chirality $\chi$ of the excited state vs the
dimensionless detuning $\delta/\Omega_0$ and phase $\tilde
\varphi_L$ of the incident laser field. b) An experimental setup
for registration of photoinduced optical rotation in the vapor
cell of H$_2$O$_2$ molecules. Two-photon Raman excitation (pump
and probe laser beams) is used for driving the tunneling
transition between left- and right-handed enantiomers. One of the
driving system laser beams (pump) is also used for partial
aligning of the molecules in the interaction region. Effect of
induced optical rotation manifests itself at the probe beam
frequency $\omega_2$ and could be measured experimentally using
polarization analyzer in the direction of pump beam.} \label{fig2}
\end{figure}

\noindent For a more general case of non-zero detuning likely
dependence is shown in Fig. \ref{fig2}a. An essential point here
is that, as one can easily see from the figure, the effect depends
on the laser field phase.

\section{An experimental scheme for registration of photoinduced
optical rotation in H$_2$O$_2$ vapor}

In this section we discuss an experiment for registration of
photoinduced optical rotation in a racemic vapor of H$_2$O$_2$
molecules.

First, we should select an effective two-level system we will work
with. It seems reasonable to use for this purpose the
$S$--$A$-transition of the first excited internal rotation level ($n=1$)
of the H$_2$O$_2$ torsional potential. Its wavelength and
frequency are $\lambda\simeq86$~$\mu$m and $\omega_{12}=
116.34$~cm$^{-1}$, respectively. The latter essentially exceeds
the corresponding frequency of 11.44~cm$^{-1}$ for the ground
internal rotation level ($n=0$).

For manipulations with the working transition we should transfer
into it some population. This can be done using two-photon Raman
excitation with two lasers frequency offset of which is equal to
the transition frequency of the $S$--$S$-transition between ground
and first excited internal rotation levels ($n=0\to n=1$). As far as
only dipole transitions are used in this excitation process, it
seems possible to almost saturate the transition in the active
volume and, therefore, for further calculations we can simply use
an estimate value of $n_S\sim1$ for the initial population value.

Due to the oscillating character of the effect, the degree of
chirality will also oscillate. Therefore, it seems most consistent
to use in experiment two laser fields (Fig.\ \ref{fig2}b). One of
them, at the frequency $\omega_1$, will serve as the pump field,
while another one with the frequency $\omega_2$ will be the probe,
with the frequency difference $\omega_1- \omega_2=\omega_{12}$ in
resonance with the tunneling transition. The effect of
photoinduced optical rotation can be registered then with the use
of two crossed polarizers, so that the registered polarization in
the direction perpendicular to the probe field polarization
manifests the effect of photoinduced optical rotation in a racemic
vapor.

In this scheme, the probe field is used also for aligning the
molecules, acting by analogy with the aligning effect in the
presence of strong dc electric field that is discussed in
Sec.~\ref{intro}. In accordance with Eq.\ (\ref{dthetas}) the
probe field affects the torsional angle, which thus oscillates
with the probe field frequency $\omega_2$. As a result, the
quadrupole Hamiltonian of the interaction with the pump field with
the frequency $\omega_1$ at the frequency $\omega_2$ and
polarization vector directed along X axis receives, in accordance
with Eq.\ (\ref{HLTLM}), a contribution at the resonance frequency
$\omega_{12}$:
\begin{eqnarray}\label{DQ}
\Delta Q&=&\frac{E_1E_2e^2a_{\rm H}^2}{8J\omega_1^2}k_L
a_{\rm O}\sqrt{1-\sin^2\vartheta \cos^2\varphi}\times\nonumber\\
&&\times\cos^2\tilde\theta\,u_1(t)u_2(t).
\end{eqnarray}

\noindent At this point, an additional dependence of $\cos(\theta
/2)$ on the torsional angle is to be included into the matrix
element (\ref{cs}), which is equivalent to the substitution
$\sin(\theta/2)\to(\sin\theta)/ 2$. The resulting matrix element
is also non-zero and additional time dependence of
$\cos(\omega_1t+ \varphi_1)$ corresponds to the substitution of
the incident laser frequency $\omega_L\to\omega_1-\omega_2$ and
the phase of the laser field $\tilde\varphi_L\to\varphi_1
-\varphi_2$. Upper bound estimation for Eq.\ (\ref{DQ}) to the order of
magnitude gives
$$ \Delta
Q\tau_p/\hbar\sim10^{-4}\sqrt{I_1I_2}\tau_p, $$

\noindent where laser pulses duration $\tau_p$ is given in seconds
and intensities $I_1,\, I_2$ of the pump and probe laser fields
are in W/cm$^2$. With this formula we can receive an upper
estimate for the required pulse intensities, which corresponds to
the action of $\pi/2$-pulse with the pulse duration of $\sim1$~ns.
The mean proportional intensity of the pump and probe pulses is
calculated then as $I_0= 10^4/\tau_p\sim
10^{13}$~W/cm$^2$.\footnote{This intensity is about the intensity
of the photoionization threshold for H$_2$O$_2$ molecules and
therefore we should keep in experiment the intensities of the pump
and probe beams on the level much less than this intensity. One
can easily fit this condition by shortening the pulse duration
(which cannot be made still much shorter than the internal
conversion characteristic time) or by increasing the sensitivity
of measuring the rotation angles. For example, for the
$\pi/20$-pulses, sensitivity of about $0.01''$ and interaction
length of about $L\sim10^{-2}$~cm the required intensities are
$I_0/100\sim10^{11}$~W/cm$^{2}$.}

The minimum value of the measured rotation angle, $\varphi_{\rm
min}$, and the expected value of specific rotation,
$\alpha=\Delta\varphi/\Delta L$, determine the length $L$ of the
active region. A rough estimate for the $\alpha$ is given by
$\alpha=k_L^2a_{\rm O} (\varepsilon-1)$, where $\varepsilon-1 \sim
10^{-4}$ is the typical dispersion value in the visible range for
H$_2$O$_2$ vapor at the normal pressure. This estimate is based on
the assumption that the order of magnitude of specific rotation is
characterized by additional small parameter $k_La_{\rm
O}\sim10^{-3}$ as compared with the linear polarization response.
The corresponding minimum length of the active region is
$L=\varphi_{\rm min}/\alpha$, which is for $\varphi_{\rm
min}=\sim1''$ yields $L\sim10^{-2}$~cm. It is important to note
here that this characteristic interaction region length is of the
same order of magnitude as the wavelength corresponding to free
precession of the tunneling transition. Therefore, the propagation
effects at this frequency are relatively small and can be
neglected.

For the minimum value of the laser beam waist diameter $w_0^2=
\lambda_L L/\pi$ \cite{demtroeder} in the active region the
corresponding pulse power is
$$
W_L=I_0w_0^2,
$$

\noindent which for the given parameters yields $10^7$ W.

An experimental setup for the discussed above two-photon Raman
excitation of the tunneling transition in the H$_2$O$_2$ vapor is
shown in Fig.~\ref{fig2}b. The effect of photoinduced optical
rotation in the racemic vapor of equally distributed enantiomers
of H$_2$O$_2$ molecule can be registered at each of the two
frequencies $\omega_{1,2}$ of the pump and probe beams. Averaging
over molecules orientation reveals only in the uncertainty of the
Rabi frequency, due to its angle dependence $\cos^2\tilde\theta$.
This, however, does not change the effect by order of magnitude in
contrast to the method of electrostatic orientation. To avoid
interference of the probe field with the frequency $\omega_2$ with
the registered at the same frequency effect in the pump field, the
beams could be slightly crossed in the sample at the angle small
enough not to affect the field distribution in the active region.

\section{Conclusions}

In conclusion, an attempt was made to answer a question: How
methods of nonlinear optics can be used to induce a required sign
of chirality in a racemic mixture of enantiomers of a chiral
molecule? We consider a racemic mixture of left- and right-handed
enantiomers of hydrogen peroxide molecule. Torsional potential of
this molecule has a characteristic double-well shape minima of
which correspond to the left- and right-handed enantiomers and are
mirror-symmetrically spaced. As a result, a splitting of
eigenstates of the torsional Hamiltonian arises from tunneling
through the potential energy barrier separating the two wells. For
H$_2$O$_2$ molecule this tunneling splitting is 11.4 cm-1 and
116.34 cm-1 for the internal rotation ground and first excited states,
respectively, so that the the molecule shows rapid oscillations
between left- and right-handed enantiomers. In a vapor or in a
solution situation is complicated due to the averaging over
ensemble.

Analysis of molecule-electromagnetic field interaction shows that
dipole interaction does not contribute to the left-right conversion
process (this interaction we can use for orientation of dipole
moments of molecules in the mixture), but the quadrupole
interaction, which leads to the excitation of coherent precession
between left- and right-handed enantiomers' states. This gives us a
tool for controlling the chiral symmetry of the molecule. We show
that biharmonic Raman excitation of the splitted internal rotation
levels can be effectively used for inducing optical activity in an
initially racemic mixture of left- and right-handed enantiomers of
H$_2$O$_2$ molecules. An experiment to study this photoinduced
optical rotation in H$_2$O$_2$ vapor is proposed.

\acknowledgments

This work was initiated by Prof. Nikolai Koroteev who devoted last
years of his life to study chiral specificity of the bioorganic
world and we dedicate it to his memory. Authors acknowledge
partial support by the Russian Foundation for Basic Research
(grant No. 96-15-96460). We thank also V.~I.~Tulin for providing
information on the isomerization potential for hydrogen peroxide
molecule and A.~Yu.~Chikishev for valuable discussion of an
experimental observation of the photoinduced chirality.

\end{document}